# First-principles study of electronic and magnetic properties of Fe atoms on $Cu_2N$/Cu(100)

Jiale Chen (陈佳乐), Jun Hu (胡军)[*]

Institute of High Pressure Physics, School of Physical Science and Technology, Ningbo University, Ningbo 315211, China.

First-principles calculations were conducted to investigate the structural, electronic and magnetic properties of single Fe atoms and Fe dimers on $Cu_2N$/Cu(100). Upon adsorption of an Fe atom onto $Cu_2N$/Cu(100), robust Fe-N bonds form, resulting in the incorporation of both single Fe atoms and Fe dimers within the surface $Cu_2N$ layer. The partial occupancy of Fe-3d orbitals lead to large spin moments on the Fe atoms. Interestingly, both single Fe atoms and Fe dimers exhibit in-plane magnetic anisotropy, with the magnetic anisotropy energy (MAE) of an Fe dimer exceeding twice that of a single Fe atom. This magnetic anisotropy can be attributed to the predominant contribution of the component along the x direction of the spin-orbital coupling Hamiltonian. Additionally, the formation of Fe-Cu dimers may further boost the magnetic anisotropy, as the energy levels of the Fe-3d orbitals are remarkably influenced by the presence of Cu atoms. Our study manifests the significance of uncovering the origin of magnetic anisotropy in engineering the magnetic properties of magnetic nanostructures.

---

[*] Email: hujun2@nbu.edu.cn

## I. Introduction

The rapid advancement of information technologies in recent years has led to the emergence of a vast amount of data, which in turn has pushed the capabilities of spintronic devices to their limits[1]. Consequently, there is a growing demand for nanometer-scale magnetic materials with specific properties to facilitate the development of high-speed, high-density, low-power, and nonvolatile spintronic devices[2,3]. To this end, single-molecule magnets (SMMs)[4,5,6] and artificial magnetic nanostructures[7,8,9,10] have been considered as potential building blocks for future spintronic devices. SMMs are typically organic molecules containing magnetic transition-metal (TM) atoms surrounded by nonmagnetic anions, which poses challenges in manipulating their magnetic properties[6]. In contrast, artificial magnetic nanostructures are composed of a few TM atoms and are minimally adsorbed onto a specific substrate[11,12,13,14,15], making them more convenient to control. Recent progresses in the atomically precise manipulation of TM atoms on substrates have demonstrated the feasibility of manufacturing atom-scale spintronic devices[16,17].

It has been revealed that TM atoms on ultrathin MgO(100) layers deposited on Ag(100) substrates, denoted as MgO(100)/Ag(100), show intriguing magnetic properties. For instance, a single Co atom on MgO(100)/Ag(100) achieved the magnetic anisotropy limit[11]; an isolated Fe atom on MgO(100)/Ag(100) manifested strong perpendicular magnetic anisotropy with the largest reported magnetic anisotropy energy (MAE) for Fe atoms adsorbed on surfaces[12]; an individual Ho atom on MgO(100)/Ag(100) exhibited a long relaxation time of magnetic remanence[13]. However, TM atoms on MgO(100) are relatively mobile, and the structural instability at high temperature hinders their potential applications in practical devices. On the other hand, nitridized Cu(100), where an insulating $Cu_2N$ layer is produced on the surface [referred to as $Cu_2N$/Cu(100)] is widely used as a substrate in experiments for exploring the magnetic properties of magnetic nanostructures. Unlike TM atoms on MgO(100)/Ag(100), a TM atom on $Cu_2N$/Cu(100) embeds in the $Cu_2N$ network, forming strong covalent bonds with N atoms[18]. Among these systems, Fe atoms on $Cu_2N$/Cu(100) have been extensively investigated either as single atoms[18,19] or atomic chains[7,20,21]. These studies have revealed that the Fe atom possesses a large spin moment ($M_S$) of about 3 $\mu_B$, and neighboring Fe atoms in an atomic chain couple with each other antiferromagnetically.

Large MAE in magnetic nanostructures is critical for preventing random spin reorientation induced by thermal fluctuations[22]. Accordingly, it is essential to uncover the origin of the magnetic anisotropy of magnetic nanostructures in both experiment and theory for engineering their magnetic properties. While the measurements of magnetic anisotropy of Fe atoms on $Cu_2N/Cu(100)$ are clear, their interpretation from a quantum viewpoint remains elusive in theory. In addition, free TM dimers, as the smallest magnetic nanostructures, usually possess large MAE[23], but they require support from a specific substrate to enable practical applications[5,24]. Therefore, investigating the magnetic properties, including magnetic anisotropy, of Fe dimers on $Cu_2N/Cu(100)$ provides an intriguing avenue of research in this field.

In this work, we conducted first-principles calculations to explore the structural, electronic and magnetic properties of Fe atoms on $Cu_2N/Cu(100)$. While a single Fe atom is embedded within the surface $Cu_2N$ layer, an Fe dimer leads to the formation of an Fe-N-Fe trimer, causing substantial structural deformations in both cases. The Fe atom, either as a single adsorbate or in an Fe dimer, exhibits large $M_S$ over 3 $\mu_B$, mainly due to the half occupation of the $d_{yz}$, $d_{xz}$ and $d_{xy}$ orbitals. Both single Fe atoms and Fe dimers prefer in-plane spin orientation along the Fe-N bonds, yielding MAEs of -0.57 and -1.66 meV, respectively. Furthermore, additional presence of Cu atoms may result in the formation of Fe-N-Cu trimers, making the MAE significantly enhanced. The origins of the MAEs in all investigated cases were revealed to be associated with contributions from the angular momentum operator $L_x$ of the spin-orbit coupling (SOC) Hamiltonian.

## II. Computational details

First-principles calculations were conducted using the Vienna *ab* initio Simulation Package (VASP)[25,26] within the framework of density functional theory (DFT). The interaction between valence electrons and ionic cores was modeled with the projector augmented wave (PAW) method[27,28]. A plane-wave basis set was employed for the expansion of electronic wavefunctions, with a kinetic energy cutoff of 500 eV. For describing the exchange-correlation potential of electrons, the generalized gradient approximation (GGA) with the Perdew–Burke–Ernzerhof (PBE) functional was utilized[29]. The SOC effect was included through a second variational procedure on a fully self-consistent basis. Convergence criteria for forces and total energy were set at 0.01 eV/Å and $10^{-5}$ eV, respectively. MAEs were calculated in the scheme of the torque method[30,31].

### III. Results and discussion

The Cu$_2$N/Cu(100) substrate was mimicked with seven Cu layers and N atoms occupying half hollow sites in the surface Cu layer, replicating the experimentally observed structure[32]. After relaxation, the Cu-N bond length within the surface Cu$_2$N network is 1.82 Å, while the bond length between N and subsurface Cu atoms is 2.19 Å. Moreover, the positions of N atoms are elevated by 0.2 Å relative to surface Cu atoms in the normal direction. These values are consistent with the experimental measurements[32]. An Fe atom tends to occupy a bridge site between two N atoms, with energy smaller by 0.36 eV and 1.64 eV than taking an atop site and hollow site, respectively, resulting in significant structural deformation, as shown in Fig. 1. Nevertheless, this deformation is localized, as evidenced by comparing structures calculated from 2×2 and 3×3 supercells, as illustrated in Fig. 1. It can be seen that the chemical bonds around the Fe atom are similar in both supercells, with deformation primarily oriented along the N-Fe-N chain and diminishing rapidly beyond the third neighbors along this direction. Therefore, a 3×3 supercell is deemed sufficiently large for investigating the electronic and magnetic properties of a single Fe atom on Cu$_2$N/Cu(100).

Figure 2(a) displays projected density of states (PDOS) of the Fe atom and its first neighboring N atom (N$_I$). It is evident that the PDOS of N$_I$ near the Fermi energy (E$_F$) is considerably smaller than that of the Fe atom, especially in the spin-down channel. Additionally, the PDOS of the Fe atom shows significant spin polarization, while the spin polarization on the N atom is negligible. Specifically, the M$_S$ on Fe amounts to 3.32 $\mu_B$, whereas a much smaller M$_S$ of 0.09 $\mu_B$ is induced on N$_I$. Furthermore, the Cu atom underneath the Fe atom also exhibits spin polarization, albeit with a small M$_S$ of 0.05 $\mu_B$. The feature of the spin polarization can be further observed from the spin density as illustrated in Fig. 2(a). It is clear that the spin density is mainly localized around the Fe atom.

The PDOS of Fe is further decomposed into its 3d orbitals, as presented in Fig. 2(b). We can see that the peak just below E$_F$ in the spin-down channel in Fig. 2(a) is predominantly contributed by the $d_{z^2}$ and $d_{x^2-y^2}$ orbitals, while the peak above E$_F$ originates mainly from the $d_{z^2}$, $d_{yz}$, $d_{xz}$ and $d_{xy}$ orbitals. Consequently, the $d_{yz}$, $d_{xz}$ and $d_{xy}$ orbitals are fully spin-polarized, each of them contributing 1 $\mu_B$ to the M$_S$, whereas the $d_{z^2}$ orbital is partially spin-polarized, contributing 0.3 $\mu_B$ to the M$_S$, similar investigation has been used for exploring the origin of spin moments of TM atoms on metallic substrate[33]. In contrast, both the spin-up and spin-down peaks of the $d_{x^2-y^2}$ orbital are below $E_F$, indicating that this orbital does not contribute to $M_S$ of Fe. This PDOS characteristic can be attributed to the features of the atomic structure. As depicted in Fig. 1, the Fe-N bond is aligned along the x direction, leading to hybridization between the N-$p_x$ and Fe-$d_{x^2-y^2}$ orbitals. Meanwhile, the Fe-$d_{z^2}$ orbital hybridizes with the $d_{z^2}$ orbital of the underlying Cu atom. These specific hybridizations not only result in a large $M_S$ on Fe but also stabilize the Fe atom on Cu$_2$N/Cu(100). Indeed, the calculated binding energy of an Fe atom adsorbed on Cu$_2$N/Cu(100) is as large as 2.73 eV.

The MAE, quantifying the energy difference between two spin orientations with the polar and azimuthal angles denoted as $(\theta_i, \phi_i)$, can be formulated as $MAE = E(\theta_1, \phi_1) - E(\theta_2, \phi_2)$. Given that Cu$_2$N/Cu(100) with an adsorbed Fe atom is actually a two-dimensional system, we first computed the MAE between in-plane and perpendicular spin orientations with respect to the surface Cu$_2$N layer as a function of azimuthal angle. As observed in Fig. 3(a), aligning the spin along the x and y directions respectively yields the lowest and highest energies of -0.57 and 1.01 meV. Hence, the easy axis is along the x direction, while the hard axis is along the y direction. On the other hand, since the Fe atom is embedded in the Cu$_2$N layer, forming a unified system, so the position of $E_F$ may influence the MAE. Accordingly, we calculated the Fermi-level-dependent MAEs in the x and y directions relative to the z direction within both 2×2 and 3×3 supercells for comparison. As shown in Fig. 3(b), the MAEs calculated in both supercells are qualitatively consistent. Moreover, the MAEs are indeed dependent on $E_F$. When $E_F$ shifts downwards to about -0.3 eV, the sign of $E_x - E_z$ changes from negative to positive, while the magnitude of $E_x - E_z$ increases when $E_F$ shifts upwards. As a consequence, the magnetic anisotropy of the Fe atom on Cu$_2$N/Cu(100) may be modulated by external electric field and gate voltage through affecting $E_F$. Note that in transition-metal nitrides and oxides, the strong correlation effect often plays a significant role. Accordingly, we evaluated this effect by adding a Hubbard U correction[34] of 2 eV and 4 eV to the 3d orbital of the Fe atom in the 2×2 supercell. As plotted in Fig. 4, the inclusion of the Hubbard U correction leads to noticeable change to the PDOS of the Fe atom. Specifically, the separation between the peaks on opposite sides of $E_F$ increases as the value of U increases and the peaks become more localized. Consequently, the value of $M_S$ on Fe increases slightly and the magnitude of magnetic anisotropy also changes, as listed in Table I. When the value of U increases, the amplitude of $E_x - E_z$ increases, while that of $E_y - E_z$ decrease. However, their signs remain unchanged. Therefore, we conclude that while the strong correlation effect may exist in our cases, it does not significantly affect the magnetic properties.

Magnetic anisotropy stems from the SOC effect, and the MAE is the consequence of the competition between in-plane and perpendicular contributions of the SOC Hamiltonian. Within the framework of the second-order perturbation theory, the MAE can be approximately expressed in terms of angular momentum operators $L_x$ (or $L_y$) and $L_z$ as[31,35]

$$MAE \approx \xi^2 \sum_{u\alpha, o\beta} (2\delta_{\alpha\beta} - 1) \left[ \frac{|\langle u\alpha|L_z|o\beta\rangle|^2}{E_{u\alpha} - E_{u\beta}} - \frac{|\langle u\alpha|L_{x/y}|o\beta\rangle|^2}{E_{u\alpha} - E_{o\beta}} \right]. \quad (1)$$

Here $\xi$ is the SOC constant; $E_{u\alpha}$ and $E_{o\beta}$ are the energy levels of an unoccupied state with spin $\alpha$ ($|u\alpha\rangle$) and an occupied state with spin $\beta$ ($|o\beta\rangle$), respectively. When the energy levels are significantly distant from $E_F$, their contributions to MAE becomes negligible due to the large denominator. As shown in Fig. 2(b), the main peaks of the spin-up PDOS are much farther away from $E_F$ compared to the spin-down peaks, implying that the MAE is dominated by the spin-down states. Accordingly, we extracted the energy levels of the primary peaks in the spin-down channel in Fig. 2(b), then individually calculated the term associated with each angular momentum operator in Eq. (1). The energy levels and nonzero terms are plotted as insets in Fig. 3(a), where the vertical lines connect pairs of energy levels with nonzero matrix elements of angular momentum operators. For $L_z$, the only nonzero matrix element is $\langle d_{x^2-y^2}|L_z|d_{xy}\rangle$, and the separation of the two energy levels ($d_{x^2-y^2}$ and $d_{xy}$) is approximately 1.4 eV, so its contribution to MAE is about $2.86\xi^2$. Similarly, there are four and two nonzero matrix elements for $L_x$ and $L_y$, respectively, contributing $-3.54\xi^2$ and $-1.28\xi^2$ to MAE, resulting in $E_x - E_z = -0.68\xi^2$ and $E_y - E_z = 1.58\xi^2$. Therefore, the energy order for the spin orientation is $E_x < E_z < E_y$, qualitatively agreeing with the result obtained from the torque method. Thus, the origin of the MAEs is successfully unveiled based on Eq. (1), wherein the angular momentum operator $L_x$ prevails over $L_z$ and $L_y$ in the contribution to MAE, thereby establishing the easy axis along the x direction.

As mentioned above, Fe dimers are anticipated to maintain significant magnetic anisotropy on $Cu_2N/Cu(100)$. Given that Fe atoms favor bridge sites, there are five possible configurations for an Fe dimer on $Cu_2N/Cu(100)$, as depicted in Fig. 5. After optimizing the atomic structures within a 5×5 supercell, we found that the configuration

'A' has the lowest energy, while the energies of the other configurations from 'B' to 'E' are higher by 0.42 eV, 1.48 eV, 1.76 eV and 0.75 eV, respectively. As shown in the bottom panel in Fig. 5, the presence of the Fe dimer induces significant structural deformation, resulting in the formation of a linear Fe-N-Fe trimer. Nevertheless, this deformation does not extend more than three neighbors beside the Fe-N-Fe trimer, similar to that induced by a single Fe atom. In addition, the bond lengths around the Fe atoms are also close to those nearby the single Fe atom on $Cu_2N/Cu(100)$, as evident from the comparison of the atomic structures in Fig. 1 and 5.

Within the configuration 'A', the two Fe atoms tend to couple with each other antiferromagnetically through superexchange interaction mediated by the N atom positioned in between, according to the Goodenough-Kanamori-Anderson rules[36,37,38]. In fact, our calculations reveal a sizable exchange energy ($\Delta E = E_{AFM} - E_{FM}$, the energy difference between the ferromagnetic and antiferromagnetic states) of -206 meV, implying strong antiferromagnetic coupling, similar to the situation in Fe chains on $Cu_2N/Cu(100)$[39]. The spin moments of the two Fe atoms are $\pm 3.13$ $\mu_B$, slightly smaller than the $M_S$ of a single Fe atom on $Cu_2N/Cu(100)$. The alteration of spin moments can be attributed to changes in the electronic structures. As presented in Fig. 6(a), the PDOS near $E_F$ is primarily dominated by the Fe atom, particularly in the spin-down channel. The sharp peak below $E_F$ in the spin-down channel is mostly contributed by the $d_{z^2}$ and $d_{x^2-y^2}$ orbitals, as shown in Fig. 6(b). Additionally, there are several peaks above $E_F$, originating from the $d_{z^2}$, $d_{yz}$, $d_{xz}$ and $d_{xy}$ orbitals. Comparing this PDOS with that of the single Fe atom on $Cu_2N/Cu(100)$ [Fig. 2(b)], we observe significant changes in the peaks of the $d_{z^2}$ orbital, with a decrease in the peak above $E_F$ and an increase in the peak below $E_F$. Therefore, the $d_{z^2}$ orbital is nearly fully occupied in this case, resulting in a notable reduction in its contribution to the $M_S$ of Fe, thus accounting for the decrease in the $M_S$ of Fe. Furthermore, the PDOS of the $d_{xy}$ orbital in the spin-down channel also undergoes significant changes, splitting into several peaks with two main peaks centered at about 0.8 eV and 1.6 eV, respectively. Evidently, this change is induced by the new structural deformation, leading to alterations in the hybridization between Fe and neighboring atoms.

The easy axis for the spin orientation of an Fe dimer on $Cu_2N/Cu(100)$ remains along the x direction, i.e. along the Fe-N-Fe chain. The MAE, $E_x - E_z$, was calculated to be -1.66 meV, with an amplitude larger than twice that of a single Fe atom on $Cu_2N/Cu(100)$. This enhancement of magnetic anisotropy is mainly ascribed to changes in the PDOS of the $d_{z^2}$ orbital as discussed earlier. As depicted in Fig. 3(a), the nonzero matrix element of $L_x$ associated with $d_{z^2}$ is $\langle d_{z^2}|L_x|d_{yz}\rangle$. The rise in the peak below $E_F$ significantly enlarges the integration of the matrix element by a factor of about 2, and its contribution to MAE is further amplified because the MAE is quadratically dependent on the matrix element as expressed in Eq. (1), resulting in significant increase in the amplitude of MAE. It is worth pointing out perpendicular magnetic anisotropy is usually desired, since the in-plane magnetic anisotropy is often weak in most cases. In other words, the energy barrier for the in-plane spin rotation is small, which ruins the stability of the spin orientation. In our cases, however, the in-plane magnetic anisotropy is large, i.e. the energy with the spin of Fe along the N-Fe-N chain (x direction) is much lower than that along other directions. Therefore, the spin orientation in our cases is specific and stable, although the in-plane magnetic anisotropy is preferred.

We note that there might be additional Cu atoms on $Cu_2N/Cu(100)$, leading to unintentional formation of Fe-Cu dimers even though a single Fe atom is desired. Therefore, we conducted calculations to investigate the electronic and magnetic properties of an Fe-Cu dimer on $Cu_2N/Cu(100)$ by substituting one Fe in the Fe dimer with a Cu atom. Interestingly, the $M_S$ of Fe remains unchanged compared to that in an Fe dimer, while the Cu atom in the Fe-Cu dimer exhibits only slight spin polarization. Nevertheless, the PDOS of the Fe atom undergoes striking changes, especially in the peak of the $d_{yz}$ orbital in the spin-down channel, as illustrated in Fig. 7(a). Therefore, the energy difference between this peak and those of the $d_{z^2}$ and $d_{x^2-y^2}$ orbitals below $E_F$ becomes considerably smaller compared to those of either a single Fe atom or an Fe dimer on $Cu_2N/Cu(100)$. As a consequence, we expect the magnitude of the MAE of an Fe-Cu dimer on $Cu_2N/Cu(100)$ to increase. As plotted in Fig. 7(b), the curve of $E_y - E_z$ near $E_F$ resembles that of the single Fe atom in Fig. 2(b). However, the curve of $E_x - E_z$ near $E_F$ exhibits notable changes in both amplitude and slope, showing a value of -2.96 meV, remarkably larger than those of both the single Fe atom and the Fe dimer. Therefore, although the Cu atom in the Fe-Cu dimer is nearly non-spin-polarized, it indeed contributes to a significant increase in the MAE of the Fe-Cu dimer by affecting the energy levels of the Fe-3d orbitals.

## IV. Conclusions

In summary, we conducted a systematic investigation of the structural, electronic and magnetic properties of Fe atoms on $Cu_2N/Cu(100)$, based on first-principles calculations. The adsorption of either a single Fe atom or an Fe dimer on $Cu_2N/Cu(100)$ induces significant structural deformation in the surface $Cu_2N$ layer, effectively embedding the Fe atom within the $Cu_2N$ layer. Notably, in the case of the Fe dimer, the two Fe atoms locate on opposite sides of the same N atom, forming an Fe-N-Fe trimer. The Fe atom in both cases possesses a large $M_S$ exceeding 3 $\mu_B$, primarily contributed by the spin polarization of the $d_{yz}$, $d_{xz}$ and $d_{xy}$ orbitals. Both single Fe atoms and Fe dimers prefer an in-plane spin orientation along the Fe-N bonds, resulting in MAEs of -0.57 and -1.66 meV, respectively. Utilizing the second-order perturbation theory, we revealed that the origin of the MAEs can be largely attributed to the coupling between the occupied $d_{z^2}$ and $d_{x^2-y^2}$ orbitals and the unoccupied $d_{yz}$ orbital through the angular momentum operator $L_x$, i.e. $\langle d_{z^2}|L_x|d_{yz}\rangle$ and $\langle d_{x^2-y^2}|L_x|d_{yz}\rangle$. Interestingly, the presence of Cu atoms may lead to the formation of Fe-Cu dimers, which can further enhance the MAE. This enhancement occurs because the energy levels of the Fe-3d orbitals are substantially influenced by the incorporation of Cu atoms.

**Acknowledgements**

This work is supported by the Program for Science and Technology Innovation Team in Zhejiang (Grant No. 2021R01004), the start-up funding of Ningbo University and Yongjiang Recruitment Project (432200942).

Table I. Spin moment ($M_s$) and magnetic anisotropy of $Cu_2N/Cu(001)$ with one Fe atom embedded in a 2×2 supercell.

| U (eV) | 0 | 2 | 4 |
|---|---|---|---|
| $M_s$ ($\mu_B$) | 3.31 | 3.51 | 3.69 |
| $E_x - E_z$ (meV) | -0.83 | -0.87 | -1.03 |
| $E_y - E_z$ (meV) | 1.10 | 0.94 | 0.63 |

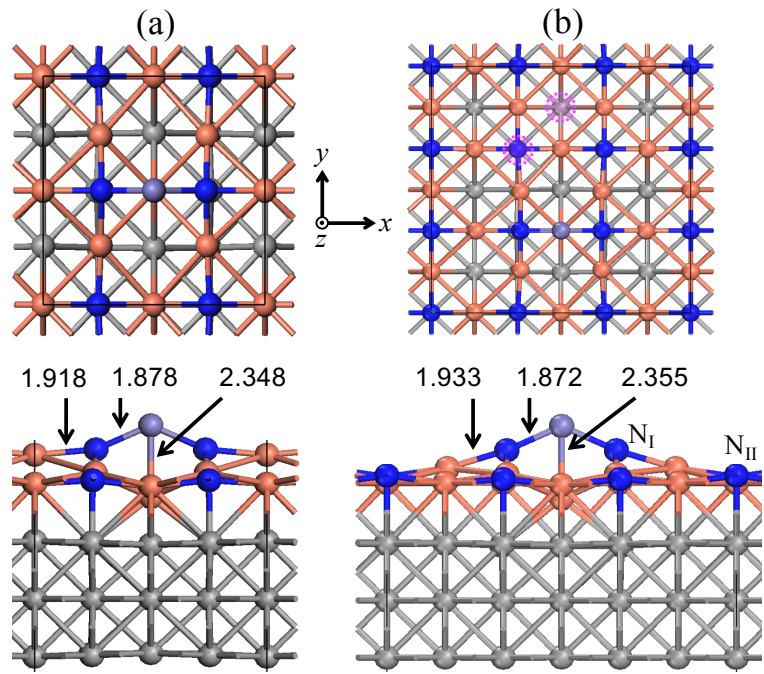

Fig. 1. Top and side views of the atomic structures with one Fe atom embedded on (a) 2×2 supercell and (b) 3×3 supercell of $Cu_2N/Cu(001)$. The grey, coral, blue, and medium purple spheres stand for bulk Cu, surface Cu, N, and Fe atoms. The purple dashed circles indicate the atop site on a N atom and the hollow site over a Cu square. Numbers denote the corresponding bond lengths in unit of Å. '$N_I$' and '$N_{II}$' stand for the first and second neighboring N atoms of the Fe atom, respectively.

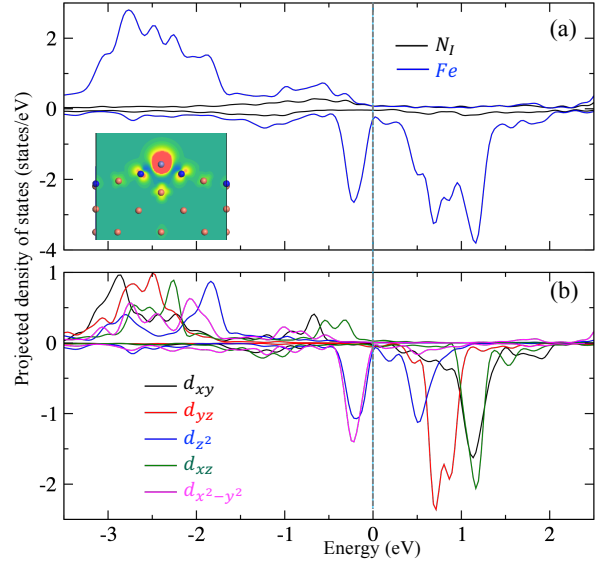

Fig. 2. Projected density of states of (a) the Fe and $N_I$ atoms and (b) the components of the Fe-3d orbital. The Fermi level is set as the zero energy. The inset in (a) displays the spin density in the plane containing the N-Fe-N chain (see Fig. 1).

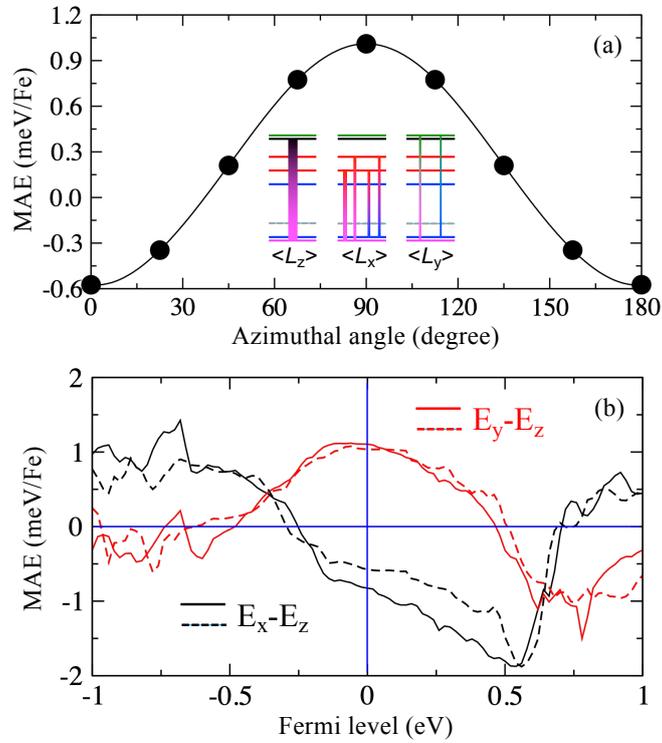

Fig. 3. Magnetic anisotropy energy (MAE) of $Cu_2N/Cu(001)$ with one Fe atom embedded. (a) In-plane MAE as a function of azimuthal angle. The energy with spin perpendicular to the surface ($E_z$) is set a zero energy. Insets display the energy levels extracted from the peaks of the PDOS in the spin-down channel in Fig. 2(b), and the nonzero terms in equation (1) presented by the vertical lines. The colors of the energy levels are the same as the corresponding orbitals in Fig. 2(b). The dashed gray lines indicate the Fermi level. The linewidths of the vertical lines denote the amplitudes of the contributions to MAE. (b) Fermi-level dependence of MAEs between x(y) direction and z direction. Solid and dashed curves are for 2×2 and 3×3 supercells, respectively.

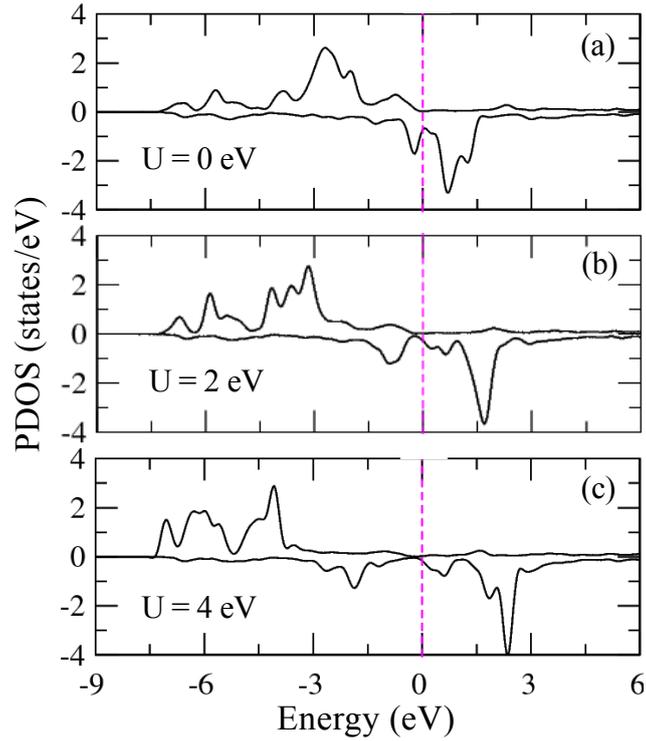

Fig. 4. Projected density of states of the Fe atom on a 2×2 $Cu_2N$/Cu(001) supercell with different values of Hubbard U term. The Fermi level is set as the zero energy.

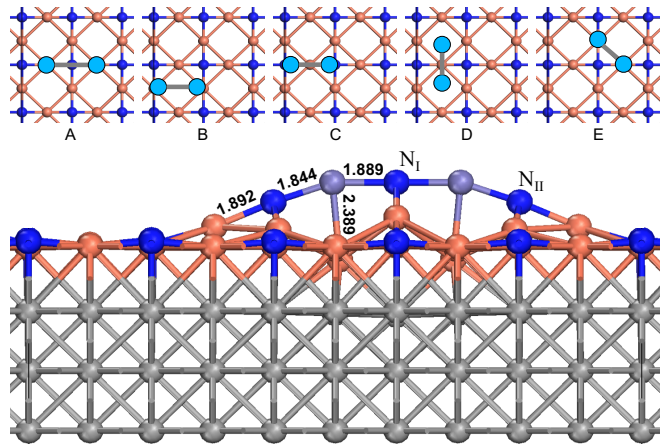

Fig. 5. Top panels: Schematic structural configurations of an Fe-Fe dimer on Cu$_2$N/Cu(001). Bottom panel: Optimized atomic structure of configuration A. The grey, coral, blue, and medium purple spheres stand for bulk Cu, surface Cu, N, and Fe atoms. 'N$_I$' denotes the N atom between the two Fe atoms, and 'N$_{II}$' stands for the first neighboring N atoms on the side of the Fe-Fe dimer. Numbers denote the corresponding bond lengths in unit of Å.

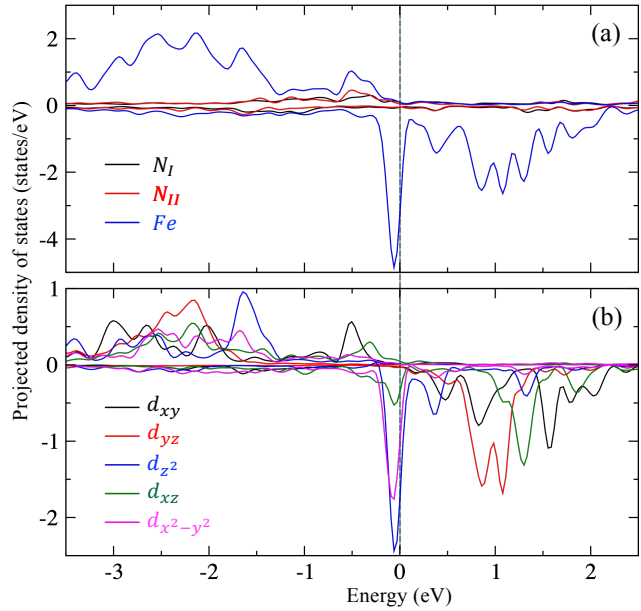

Fig. 6. Projected density of states of (a) the Fe, N$_I$ and N$_{II}$ atoms and (b) the components of the Fe-3d orbitals for the Fe-dimer on Cu$_2$N/Cu(001). The Fermi level is set as the zero energy.

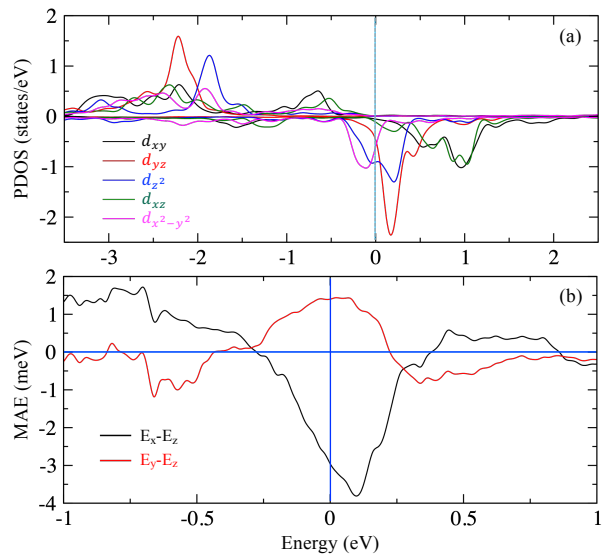

Fig. 7. (a) Projected density of states (PDOS) of the Fe-3d orbitals in the Fe-Cu-dimer on Cu$_2$N/Cu(001). (b) Fermi-level-dependent magnetic anisotropy energy between x(y) direction and z direction. The Fermi level is set as the zero energy.